\documentstyle[amssymb,prb,twocolumn,aps,epsfig]{revtex}

\begin{document}

\twocolumn[\hsize\textwidth\columnwidth\hsize\csname
@twocolumnfalse\endcsname
\title{Localized $f$ electrons in Ce$_x$La$_{1-x}$RhIn$_5$: dHvA
Measurements}
\author{U.~Alver$^1$, R.~G.~Goodrich$^1$,
N.~Harrison$^2$, Donavan Hall$^3$, E.~C.~Palm$^3$, T.~P.~Murphy$^3$,
S.~W.~Tozer$^3$,
P.~G.~Pagliuso$^4$, N.~O.~Moreno$^4$, J.~L.~Sarrao$^4$ and Z.~Fisk$^3$}
\address{$^1$Department of Physics and Astronomy,
Louisiana State University, Baton Rouge, Louisiana 70803\\
$^2$National High Magnetic Field
Laboratory, LANL, Los Alamos, Mew Mexico 87545\\
$^3$National High Magnetic Field
Laboratory, Florida State University, Tallahassee, Florida 32310\\
$^4$Los Alamos National Laboratory, Los
Alamos, New Mexico 87545}

\date{\today}
\maketitle

\begin{abstract}
Measurements of the de Haas-van Alphen effect in
Ce$_{x}$La$_{1-x}$RhIn$_{5}$ reveal that the Ce $4f$ electrons remain
localized for all $x$, with the mass enhancement and progressive loss
of one spin from the de
Haas-van Alphen signal resulting from spin fluctuation effects.
This behavior may be typical of antiferromagnetic
heavy fermion compounds, inspite of the fact that the $4f$
electron localization in CeRhIn$_5$ is driven, in part, by a spin-density
wave instability.
\end{abstract}
]\narrowtext

Ce and U-based heavy fermion systems display a variety of different
ground states,
ranging from ordered antiferromagnets to superconductors
\cite{stewart1}. There has
been considerable effort made to correlate these changes in behavior
with changes in the electronic structure, and to identify those
characteristics that are universal to all heavy fermion systems
\cite{hewson1}. With this
goal in mind, de Haas-van Alphen (dHvA) experiments continue to play
an important role \cite{harrison1,haga1,goodrich1,aoki1}.
They have the potential to yield otherwise
inaccessible information on how the $4f$ or $5f$ electrons mix with the
conduction electrons, and, ultimately, how this determines the
electronic ground state.

The heavy fermion compound CeRhIn$_5$ is of
special interest to the general understanding heavy fermion systems because,
not only does it order
antiferromagnetically below $T_{\rm N}\sim$~3.8~K, but also becomes
superconducting ($T_{\rm c}\sim$~2.1~K) at hydrostatic pressures $p$
exceeding $p_{\rm c}\sim$~16 kbar \cite{hegger1}. Thus, by changing
the experimental
conditions, CeRhIn$_5$ displays the two most contrasting types of behavior
that are found in heavy fermion systems \cite{stewart1}.
A matter of further interest is that both the
electronic contribution to the specific heat $\gamma T$ and $T_{\rm N}$
remain largely unchanged for pressures $p<p_{\rm c}$ \cite{hegger1},
and under the application of high magnetic fields $B\leq$~50~T,
\cite{takeuchi1,Kim}. In addition, the value of the critical applied field
that changes or destroys the antiferromagnetic state is
highly anisotropic, ranging from 2 T for $B\|[100]$ to over 50 T for
$B\|[001]$ \cite{takeuchi1}.  The 2 T transition for $B\|[001]$ is
known to be from
one spiral spin state to a second that persists to $\sim$~50~T 
\cite{takeuchi1}.

In this paper, it is shown that the apparent robustness of the electronic
structure to large variations of $p$ and $B$ can be attributed to the
nearly ideal
localized behavior of the $4f$ electrons in CeRhIn$_5$. dHvA
experiments performed throughout the entire Ce$_{x}$La$_{1-x}$RhIn$_{5}$
series, (i.e. for 0~$<x<$~1) reveal that the Fermi
surface topology is nearly independent of $x$, implying
that there is virtually no exchange of charge degrees of freedom between
the $4f$ and conduction electrons. This type of behavior may typify
antiferromagnetic heavy fermion systems. The
progressive loss of one spin from the dHvA signal together with an
increase in the quasiparticle
effective masses on increasing $x$ indicate that spin
fluctuation effects are nevertheless important.

Single crystal samples of Ce$_x$La$_{1-x}$Rh In$_5$, with
0~$\leq x \leq$~1 in 12 incremental steps, were grown in In flux
\cite{hegger1}.
They were
etched in a 25\% aqueous HCl solution both to remove the
residual flux and to reduce them to a size necessary for dHvA
measurements. A variety of experimental techniques were used in order
to cover a wide range of frequencies and magnetic fields, $B$. Torque
magnetometry and field modulation measurements were made in static
magnetic fields at the National High Magnetic Field
Laboratory (NHMFL), Tallahassee, while
pulsed magnetic field measurements in fields of up 50~T were carried
out at the NHMFL, Los Alamos (detailed discussions of the techniques
are contained in References \cite{harrison1,goodrich1}). Temperatures in the
range 0.45~$\leq T \leq$~6 K were used
in the measurements. All three techniques yielded
similar frequencies for $B$ applied
along the principal [100] and [001] axes that were under investigation.

Examples of Fourier transforms of the dHvA data (in the $1/B$ domain) for
$x=$~0.05 and $B$
applied along [100] and for $x=$~0.5 and $B$ applied along [001] are
shown in Figs.\ref{dHvA}(a) and
\ref{dHvA}(b) respectively. The
presence of many frequencies with signal-to-noise ratios comparable to those
measured in
pure LaRhIn$_5$ \cite{hall1} and CeRhIn$_5$ \cite{cornhole1,hall2}
shows that no significant
attenuation of the dHvA amplitude results from alloying. This finding
is similar to that observed throughout the Ce$_x$La$_{1-x}$B$_6$ series.
In contrast to Ce$_x$La$_{1-x}$B$_6$ \cite{goodrich1}, however, measurements on
Ce$_x$La$_{1-x}$RhIn$_5$ (for $0\leq x\leq 1$), shown in Fig.
\ref{frequencies}, reveal no significant dependence of the
Fermi surface topology on $x$, apart from two small exceptions:
First, one very low frequency in LaRhIn$_5$ of $\sim$~7~T,
originating from a pocket that
occupies less than 1 part in 10$^4$ of the Brillouin zone volume, is
not present
in the alloys \cite{goodrich2}.
Second, the two distinct but closely spaced frequencies,
f$_6$ and f$_7$ (for $B\|[001]$) only can be individually resolved for
$x\leq$~0.95.

In pure CeRhIn$_5$,
there has been some uncertaintly about the degree to which the $4f$ electrons
contribute to the Fermi surface \cite{cornhole1,hall2,harima1}. Because
the bandstructure
calculations on pure CeRhIn$_5$ consider the $4f$ electrons to be
fully itinerant, only if the $4f$ electron wavefunctions are
well mixed with the
conduction electrons might one expect to find good agreement
between dHvA data and the calculations.
Some degree of similarity between the dHvA data and the itinerant $4f$ electron
bandstructure was reported in CeRhIn$_5$ \cite{hall2}.
However, the average fractional frequency difference between the
calculated and measured values, (F$_{Exp}$-F$_{BS}$/F$_{Exp}$), is
0.44 for $B\|$[001].  Therefore, it is unclear from the
measurements and calculations in Ref. \cite{hall2} whether the $4f$
electrons are itinerant or localized.
The results presented in Fig. \ref{frequencies} show that, because of the
absence of any strong dependence of the frequencies on $x$, the
addition of $4f$
electrons does not change the Fermi surface volume and therefore they must
be almost entirely localized.
This conclusion is rigorous since it does not depend on the
ability of bandstructure calculations to reproduce experimental data.

A number of Ce-based heavy fermion systems, for example CeB$_6$,
CeRu$_2$Si$_2$, CeAl$_2$ and CeCu$_2$Si$_2$,
have been shown to exhibit dHvA oscillations at high magnetic fields that
appear to be consistent with a localized $4f$ electron picture
\cite{harrison1,aoki1,springford1,hunt1}. By localized we mean that the
high magnetic field dHvA data in these systems are more closely
reproduced by bandstructure calculations made on the La analogue compounds
than on the completely delocalized  bandstructure
calculations for the Ce compounds. It is at high
magnetic fields, where $f$ electron alignment occurs in these
materials, that the dHvA
data are often most clear, thereby facilitating a direct
comparison between experiment and theory. In cases where this has
been studied, it has been concluded
previously\cite{harrison1,aoki1,springford1,hunt1} that
{\it all} Ce-based heavy fermion compounds exhibit localized $f$ electron
behavior upon their alignment by a magnetic field. Owing
to the absence of a simple equivalent analogue to La in the case of U,
it is difficult to verify in this manner whether this is true in U-based
heavy fermion systems.

Several heavy fermion compounds now have been shown to
undergo Fermi surface changes at, what is commonly referred to
as, the metamagnetic transition field $B_{\rm M}$.
In CeRu$_2$Si$_2$, this change has been associated with what appears
to be a transition from itinerant
to localized $4f$ electron behavior at the field where the $4f$
electrons become aligned \cite{aoki1}. The Fermi
surface change in CeCu$_2$Si$_2$ also is likely to be consistent with
this picture \cite{hunt1}. In CeIrIn$_5$, on the other hand,
dHvA experiments only have been performed at fields below
$B_{\rm M}$ ($\sim$~40~T) \cite{takeuchi1},
where the dHvA results for the fractional frequency differences
between measured and fully itinerate calculated band
structure\cite{haga1} frequencies
are only 0.15 compared to the 0.44 in CeRhIn$_5$. In addition, at
least three U-based
heavy fermion systems (UPd$_2$Al$_3$, UPt$_3$ and URu$_2$Si$_2$), in
which Fermi
surface transformations are observed, have been shown to
possess Fermi surfaces at low magnetic fields consistent with
the itinerant $5f$ electron bandstructure \cite{terashima1}.

In this context, CeRhIn$_5$ is interesting because it exhibits localized $4f$
electron behavior at {\it all} magnetic fields over which dHvA
oscillations are observed, from fields as low as 4~T up to 50 T over a
range in which the magnetization never saturates \cite{takeuchi1}. In
this respect,
CeRhIn$_5$ exhibits qualitatively the same behavior as CeAl$_2$;
another well known
strictly antiferromagnetic heavy fermion compound in which the $4f$ electrons
are localized at fields both above and below a magnetic transition
\cite{springford1}. Because their Fermi surfaces do not change on
crossing magnetic transitions, the
antiferromagnetic heavy fermion systems CePb$_3$ \cite{ebihara1} and
UCd$_{11}$ \cite{cornhole2} also may exhibit the
same behavior.
The pattern of behavior that emerges here is that a simple
almost-localized $4f$ electron model appears to apply very well to
those systems in which the Kondo screening is weak and in which
Ruderman-Kittel-Kasuya-Yosida (RKKY) interactions dominate to
form a strictly antiferromagnetically ordered ground state; namely CeRhIn$_5$,
CeAl$_2$, CePb$_3$, CeB$_6$ and UCd$_{11}$, although dHvA studies
only have been possible in CeB$_6$ at fields that greatly exceed $B_{\rm
M}$ \cite{harrison2}.

Given that bandstructure calculations on CeRhIn$_5$ predict most of
the $4f$ electron spectral weight to exist at energies between 0.3 and 0.7~eV
above the Fermi energy $\varepsilon_{\rm F}$ \cite{hall2}, the
on-site correlation
energy $U_{ff}$ (which is ultimately responsible for their localized
behavior), must greatly exceed 0.7~eV. In order for the $4f$
electrons not to modify the Fermi surface topology in
Ce$_x$La$_{1-x}$RhIn$_5$, the singly occupied core $4f$ electron
level $\varepsilon_{\rm 4f}$ must also be lower than $\varepsilon_{\rm F}$
by an amount that significantly exceeds the hybridization energy $V$.

In spite of the fact that the $4f$
electrons are almost entirely localized in CeRhIn$_5$, the effects of spin
fluctuations (excitations involving only spin degrees of freedom) are clearly
important. This is seen in Fig. \ref{masses} by the
progressive increase in the
effective mass across the series and in Fig.
\ref{spin}(a) by the progressive loss of one spin direction from the
dHvA signal.
A complete single spin dHvA signal is observed for x $\geq$ 0.4, when the
logarithm $\ln[A_q/\sqrt{q}]$ of the amplitude $A_q$ of each dHvA harmonic
divided by the square
root of the harmonic index $q$ falls on a straight line. We note that
this is  near the concentration of Ce where the series first exhibts
antiferromagnetism.
This type of behavior recently was observed in Ce$_x$La$_{1-x}$B$_6$
and the sign of the dominant spin channel was further found to be
consistent with
the model of Edwards and Green \cite{goodrich1,edwards1}.
According to their model, the high
magnetic field phase of a heavy fermion system is similar to that of a
field-induced ferromagnet.
The spin-down electrons have effective masses that
are more greatly enhanced by spin fluctuations and suffer from
disorder scattering,
causing them to no longer contribute to the dHvA signal
\cite{goodrich1,edwards1}.  In the present case, the $4f$
moments are only partially alligned in the range of measurement fields used.

The variation in the effective mass $m^\ast$ on increasing $x$ from the single
impurity limit ($x\rightarrow$~0) to the impurity lattice limit
($x=$~1), has yet to be understood \cite{hewson1}. The model of Gor`kov and Kim
\cite{gorkov1}, which considers two-impurity terms, appeared to
explain a quadratic dependence in Ce$_x$La$_{1-x}$B$_6$ over a
limited range of $x$. This model cannot, however, account for the
non-monotonic dependence of $m^\ast$ on $x$ observed in
Ce$_x$La$_{1-x}$RhIn$_5$
displayed in Fig. \ref{masses}.
Were the magnetic interactions between
the $4f$ electrons only of the nearest neighbor RKKY type, one
would normally expect the establishment of long range
antiferromagnetic order in pure CeRhIn$_5$ to compete with the
Kondo-like screening interactions that give rise to strongly enhanced effective
masses \cite{hewson1}. It can be seen that there is only a slight
increase in $m^\ast$ near $x=$~0.4 where the onset of antiferromagnetism
occurs in the series.  The dramatic increase in $m^\ast$ for
$x\geq$~0.9 suggests that additional terms need to be incorporated
into the lattice model.

Neutron diffraction experiments
recently have shown that the periodicity of the magnetic structure
perpendicular to the CeIn$_3$ planes is incommensurate, evidencing the
existence of a spin-density wave (SDW) \cite{bao1}.
This could imply that, in
addition to RKKY interactions,
Fermi surface nesting plays an important role in causing the
$4f$ electrons to behave in a localized fashion. While we would
expect RKKY interactions
to be especially important within the CeIn$_3$ planes \cite{bao1},
Fermi surface nesting could help stabilise long range order
perpendicular to the planes.
One particularly novel aspect of this observation is that the
localised spins that accompany
SDW ground states invariably result from electrons that are predisposed to
be itinerant prior to their condensation \cite{gruner1}.
While the section of Fermi surface responsible for nesting has yet to be
identified, the collapse of the f$_6$ and f$_7$ frequencies into a
single frequency in pure CeRhIn$_5$ suggests that the Fermi surface
becomes more two-dimensional in the limit $x\rightarrow$~1; increasing
the likelihood of nesting \cite{gruner1}. The possible involvement of
Fermi surface
nesting
indicates that the balance between itinerant and
localized $4f$ electron nature in CeRhIn$_5$ may be somewhat delicate.

In summary, on performing a dHvA study across the
Ce$_x$La$_{1-x}$RhIn$_5$ series, the near insensitivity of the Fermi
surface topology to $x$ implies that the $4f$ electrons
are almost entirely localized. The increase in the effective masses
together with the
disappearance of one spin contribution to the dHvA signal on
increasing $x$ do, however, indicate that spin fluctuation effects are
important. The virtual insensitivity of $T_{\rm N}$ and $\gamma$ to $p$, for
$p<p_{\rm c}$ \cite{hegger1}, implies that the localised $4f$ electron picture
survives for $p<p_{\rm c}$, but then gives way to a more abrupt Fermi surface
change for $p>p_{\rm c}$ accompanying the onset of superconductivity.
Such a Fermi surface change could be consistent with the finding, in other
superconducting heavy fermion systems (CeCu$_2$Si$_2$
\cite{springford1},
UPd$_2$Al$_3$, UPt$_3$,
URu$_2$Si$_2$ \cite{terashima1} and CeIrIn$_5$ \cite{haga1}), where the dHvA
data for $B<B_{\rm M}$ are more consistent with an itinerant $f$ electron
bandstructure.

Work conducted at the National High Magnetic Field Laboratory was
supported by the National Science Foundation (NSF), the State of
Florida and the Department of Energy. One of us (ZF) acknowledges
support under NSF grant DMR-9971348.

\begin{center}
\begin{figure}
    \epsfig{file=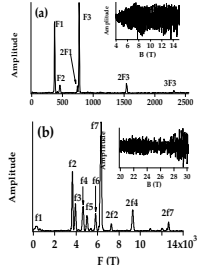, height=.42\textwidth}
\caption{Fourier transform of dHvA signals in Ce$_x$La$_{1-x}$RhIn$_5$,
(a) for 4~$<B<$~14.5~T aligned along the [100]
axis at $T\approx$~1.30~K and $x=$~0.05, and (b) for 20~$<B<$~30~T
aligned along
the [001] axis at $T\approx$~1.43~K and $x=$~0.5. The frequency
labels are the same
as in [11] and the data are shown in the insets.}
\label{dHvA}
\end{figure}

\begin{figure}
    \epsfig{file=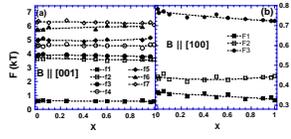, height=.42\textwidth}
\caption{The principal dHvA frequencies in Ce$_x$La$_{1-x}$RhIn$_5$
plotted versus $x$, (a) for $B$ applied along [100] and (b) for $B$
applied along [001].}
\label{frequencies}
\end{figure}

\begin{figure}
    \epsfig{file=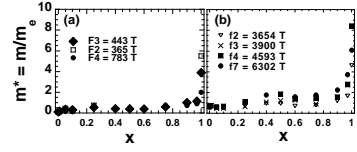, height=.42\textwidth}
\caption{Concentration dependence of the effective masses of various
dHvA frequencies for (a) 4~$<B<$~14.5~T aligned along [100] and (b)
20~$<B<$~30~T aligned
along [001]. The effective masses are obtained by fitting only to the
temperature dependent term in the Lifshitz-Kosevich formula.}
\label{masses}
\end{figure}

\begin{figure}
    \epsfig{file=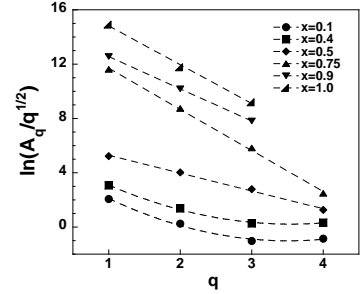, height=.42\textwidth}
\caption{A plot of $\ln[A_q/\sqrt{q}]$ versus harmonic index $q$ for
the F$_3$ frequency, for which several harmonics could be observed.
The dHvA signal is dominated by a single spin
for $x\geq$~0.5, evidenced by the straight lines.}
\label{spin}
\end{figure}
\end{center}

\end{document}